\documentclass[journal]{vgtc}                     

\onlineid{0}

\vgtccategory{Research}

\title{Prediction market visualizations, betting, and uncertainty: A study of Reddit Posts and Comments}

\author{%
  \authororcid{Subham Sah$^\pi$}{0009-0007-9446-6731},
  \authororcid{Alireza Karduni$^\pi$}{0000-0001-9719-7513},
  \authororcid{Douglas Markant}{0000-0003-0568-2648},
  \authororcid{Wenwen Dou}{0000-0003-0319-9484}
}

\authorfooter{
 \item Subham Sah, Douglas Markant, and Wenwen Dou are with the University of North Carolina at Charlotte. Email: \{ssah1, dmarkant, wdou1\}@charlotte.edu.
  \item
  	Alireza Karduni is with Simon Fraser University.
  	E-mail: akarduni@sfu.ca.
    
$^\pi$ = authors contributed equally
}

\abstract{%
Prediction market platforms present contracts about future events through visualizations that show probabilities, prices, trends, odds, and payout information. Although these visualizations often appear precise, they do not always show uncertainty directly. As a result, users infer uncertainty from market movement, visualization cues, and contextual information. In this paper, we examine how users interpret prediction market visualizations through a qualitative analysis of posts and comments from the Reddit community r/Kalshi. From an initial corpus of approximately 12,000 posts and 96,000 comments, we identified 360 posts containing prediction market visualizations and conducted a thematic analysis of annotated posts and related discussions. Our findings show that users infer uncertainty through several forms of interpretation: they interpret chart values, struggle with probability information displayed, bring in external knowledge, question credibility and liquidity, critique visualization design, and connecting visualized information to betting decisions. 
}


\teaser{
  \centering
  \includegraphics[width=\linewidth, alt={A view of a city with buildings peeking out of the clouds.}]{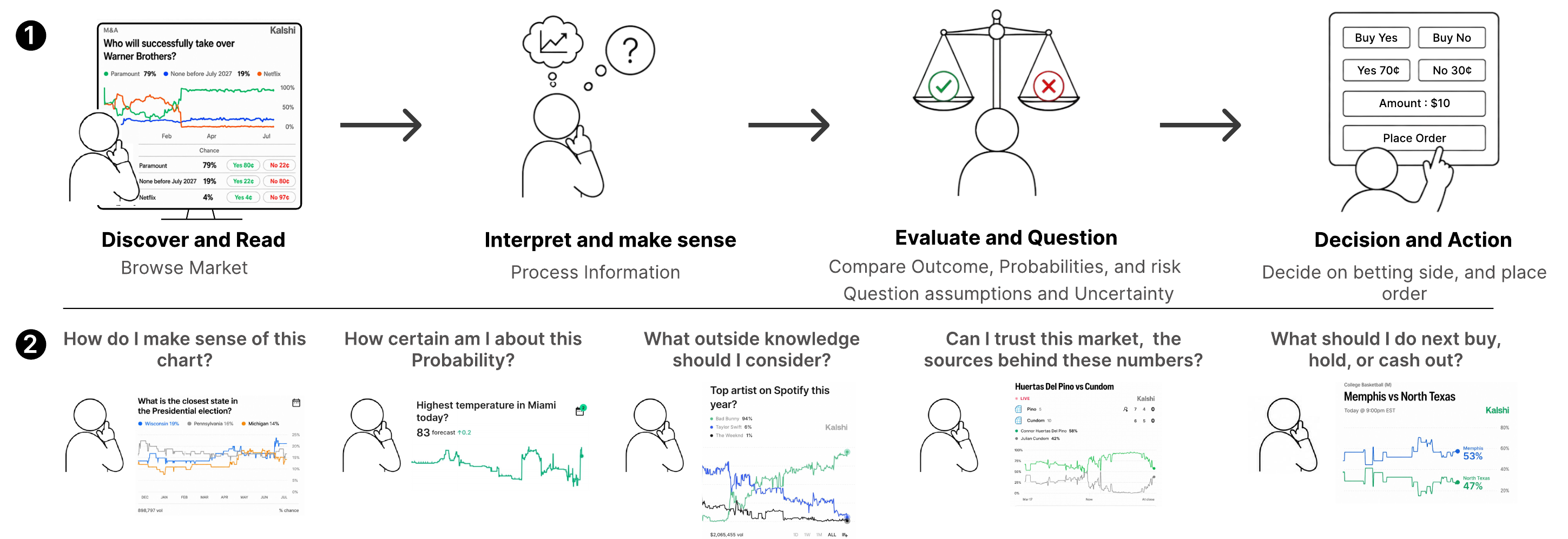}
  \caption{%
  	Overview of the study (\textbf{1}) A typical user journey through prediction market on Kalshi, (\textbf{2}) Themes identified through our qualitative analysis.
  }
  \label{fig:teaser}
}

\graphicspath{{figs/}{figures/}{pictures/}{images/}{./}} 

\usepackage{tabu}                      
\usepackage{booktabs}                  
\usepackage{lipsum}                    
\usepackage{mwe}                       
\usepackage{ccicons}                   

\usepackage{mathptmx}                  
\usepackage{booktabs}
\usepackage{tabularx}
\usepackage{multirow}
\usepackage{array}
\usepackage{adjustbox}
\usepackage{graphicx}

\graphicspath{{figs/subsection-icons/}}

\begin{document}

\raggedbottom
\setlength{\textfloatsep}{8pt}
\setlength{\floatsep}{8pt}
\setlength{\intextsep}{8pt}


\firstsection{Introduction and Background}

\maketitle

Prediction market platforms (e.g., Polymarket \cite{polymarket2026}, Kalshi\cite{kalshi2026}, Robinhood\cite{robinhood2026}) have recently surged in popularity, allowing users to bet real money on the probabilistic outcomes of future events ranging from election results to AI capability growth and economic indicators. Prediction market visualizations present a distinctive challenge for uncertainty visualization. Although these platforms often display precise-looking probabilities, prices, and trends, the underlying quantities correspond to uncertain future events and financially consequential decisions. Recent reporting and empirical analyses suggest that many casual participants lose money on prediction market platforms, while profits are concentrated among more sophisticated traders \cite{mehta2026almost,predictionmarkets2026whowins}. This raises an important visualization question: how do users infer uncertainty when charts do not explicitly encode it, but instead embed uncertainty in market movement, outcome categories, odds, payouts, and platform-specific visualization cues?

 This is a challenging issue in many financial allocation and forecasting tasks, which are strongly influenced by how uncertainty and risk are represented. Padilla et al.'s cognitive framework for decision-making with visualizations demonstrates that judgments are often informed by perceptual and heuristic processes in addition to analytical reasoning \cite{padilla2018decision}. Prior research in uncertainty visualization further shows that different visual representations of probabilistic information can influence perceptions of reliability, confidence, and risk, ultimately affecting decision-making outcomes \cite{hullman2019authors,correll2014error,hullman2015hypothetical}. Wesslen et al. further show that uncertainty visualization can affect investment decision, including how users reason about risk, losses and long-term investment returns \cite{wesslen2021effect}. 
 
 Prediction markets provide a particularly relevant setting for examining these processes because interpretation is tied to financial risk and participants frequently interpret probability-over-time charts, forecast trajectories, market trends, and payout information when evaluating potential outcomes. As information aggregation mechanisms, these markets depend on participants correctly interpreting probabilistic signals and updating their beliefs as conditions change \cite{wolfers2004prediction,berg2008results}. While prior prediction market research has primarily focused on forecasting accuracy, information aggregation, and market efficiency \cite{wolfers2004prediction,berg2008results,manski2006interpreting}, few studies have examined how users interpret and discuss the visual representations through which these predictions are communicated. Understanding this user interpretation is important because recent reporting and empirical analyses suggest that losses are common among participants, while profits are concentrated among more trained or seasoned traders \cite{mehta2026almost,predictionmarkets2026whowins}.

The need to understand user interpretation and uncertainty reasoning is important in settings where uncertainty is communicated through visual forms rather than explicitly encoding uncertainty. Prior studies in uncertainty visualization demonstrate that visual encodings such as error bars, confidence intervals, distributional displays, and hypothetical outcome plots influence how viewers reason about reliability, variability, and risk \cite{correll2014error,hullman2015hypothetical,kale2018hypothetical}. These findings highlight that different representations of uncertainty can lead to different conclusions about the same underlying data, underscoring the importance of understanding how uncertainty is perceived and interpreted by viewers. Research on visualization literacy further emphasizes that successful interpretation depends on more than the visual encoding itself. Users bring prior knowledge, expectations, domain expertise, and reasoning strategies that influence how visual information is understood. Boy et al. introduced one of the earliest systematic frameworks for assessing visualization literacy \cite{boy2014principled}, while later studies also examined how people evaluate visual evidence and recognize misleading visualizations \cite{ge2023calvi}. These considerations are particularly relevant in prediction market environments, where probabilistic information is frequently presented through dynamic and interactive visualization.

The data visualization interpretation challenges are especially visible in online discussions, where users explain, question, and debate what visualizations mean. Social media platforms and online communities offer a valuable setting for studying how people interpret, debate, and critique prediction market visualizations. Prior work shows that visualizations can shape beliefs, influence judgments, and credibility of information \cite{pandey2014persuasive,pandey2015deceptive}. Public discussions around visualizations can also reveal how users evaluate evidence, question interpretations, and make sense of visualized data \cite{lisnic2024yeah,kauer2021public}.
To study how users interpret prediction market visualizations, we examined Reddit discussion form r/Kalshi subreddit community. Through qualitative analysis of posts and comments' discussions around charts, probabilities, forecast charts, market movements, and other visualization features, we analyzed how users interpret uncertainty, evaluate visual evidence, and make sense of probabilistic information in real world online discourse.

The \textbf{research question} in this study we primarily ask : \textit{How do users infer uncertainty from prediction market visualizations and associated market-interface cues? 
} 


\section{Data Collection and Methodology}
We collect data from r/Kalshi, a reddit \cite{reddit2026} community focused on prediction markets, forecasting, and trading strategy discussions. Using the Reddit API, we pulled roughly 12,000 posts and 96,000 comments, along with post metadata, titles, text content, comment threads, and any linked images or screenshots.

\subsection{Visualization-Based Filtering}

To support the study’s focus on how users interpret and discuss prediction market visualizations, we needed to narrow this corpus down to posts that actually contained visual content. We first pulled posts with images, screenshots, or externally linked visual media, which left about 5,600 posts.

To further identify posts that contain data visualizations, we employed a vision-language model \cite{hui2024qwen2} to analyze the visual content associated with each post. The model was used to detect the presence of visualization types relevant to prediction markets, including line charts, multi-line charts, forecast charts, probability-over-time charts, market trend charts, candlestick charts, bar charts, odds or payout charts, and screenshots of trading interfaces containing embedded charts. Posts containing photographs, memes, news screenshots without charts, promotional graphics, text-only images, or other non-visualization content were excluded from further analysis. This process resulted in a final dataset of 360 posts that contain prediction market visualizations and their associated discussion threads.

After filtering, the final dataset contained 360 posts with prediction market visualizations and their associated comments. The dataset included charts, market screenshots, probability displays, and other visual elements through which users encountered and discussed uncertain future outcomes.

\section{Insights from Reddit Posts and Comments}

Building on the final dataset of 360 posts, we conducted a thematic analysis of the visualizations, post content, and associated comments to understand how users inferred uncertainty from prediction market information.
We focused on comments where users' struggled to understand chart values, questioned the meaning of probabilities or payouts, used outside information to explain market movement, raised concerns about credibility or design, or connected visual patterns to betting decisions.

During the analysis process, we created annotations that captured recurring forms of interpretation and response. These included confusion about multiple bet outcomes, uncertainty about whether a chart represented a forecast or a market price, difficulty interpreting probabilities, attempts to understand chart values through tooltips, comments about probabilities not adding up to 100\%, references to outside knowledge, skepticism toward data accuracy, concerns about mislabeling, visualization design issues, technical problems, and behavior or emotion oriented reactions. These annotations were then grouped into broader themes that describe how users made sense of uncertainty in prediction market visualizations. 

\subsection{Findings}

Our thematic analysis revealed different themes relating to data interpretation and chart sensemaking, uncertainty and probability comprehension, external knowledge and contextual reasoning, skepticism and credibility on the data and visualization, visualization design and it's usability. Table~\ref{tab:themes-codes-prevalence} summarizes these themes, their associated codes, and their prevalence counts. We organize the findings around these themes and the broader questions that emerged from the analysis.

\newcolumntype{Y}{>{\raggedright\arraybackslash}X}
\newlength{\tabwidth}
\setlength{\tabwidth}{\linewidth}

\begin{table}[t]
\centering
\scriptsize
\renewcommand{\arraystretch}{1.1}
\setlength{\tabcolsep}{4pt}

\begin{tabularx}{\tabwidth}{
    p{0.42\tabwidth}
    X
    p{0.12\tabwidth}
}
\toprule
\textbf{Theme} & \textbf{Code} & \textbf{Prevalence} \\
\midrule

\multirow{3}{=}{\textbf{Data Interpretation \& Chart Sensemaking} \scriptsize (n=35)}
& Chart value interpretation & 32 \\
& Market mechanics \& payouts & 10 \\
& Tooltips \& data-source reading & 4 \\
\midrule

\multirow{3}{=}{\textbf{Uncertainty \& Probability Comprehension} \scriptsize (n=11)}
& Multiple outcomes \& hedging & 3 \\
& Probability aggregation/mismatch & 7 \\
& Forecast vs. market price & 2 \\
\midrule

\multirow{2}{=}{\textbf{External Knowledge \& Contextual Reasoning} \scriptsize (n=22)}
& Outside event/domain knowledge & 18 \\
& External sources \& resolution rules & 4 \\
\midrule

\multirow{2}{=}{\textbf{Visualization Design \& Usability} \scriptsize (n=9)}
& Visualization clarity, layout, and color & 7 \\
& Misleading labels/display mismatch & 3 \\
\midrule

\textbf{Skepticism \& Credibility Assessment} \scriptsize (n=11)
& Liquidity, trust, and source credibility & 11 \\
\midrule

\multirow{2}{=}{\textbf{Engagement \& Decision-Oriented Reactions} \scriptsize (n=17)}
& Advice seeking \& betting strategy & 7 \\
& Emotional reactions \& conviction & 10 \\

\bottomrule
\end{tabularx}

\caption{Preliminary themes, codes, and prevalence 
(\(n=66\) annotated snippets; codes are non-mutually exclusive).}
\label{tab:themes-codes-prevalence}
\end{table}

\subsection{\textbf{Are these number probabilities, are there enough information? : Data Interpretation and Chart sensemaking}}

\begin{figure}[t]

  \centering
  \includegraphics[width=\linewidth]{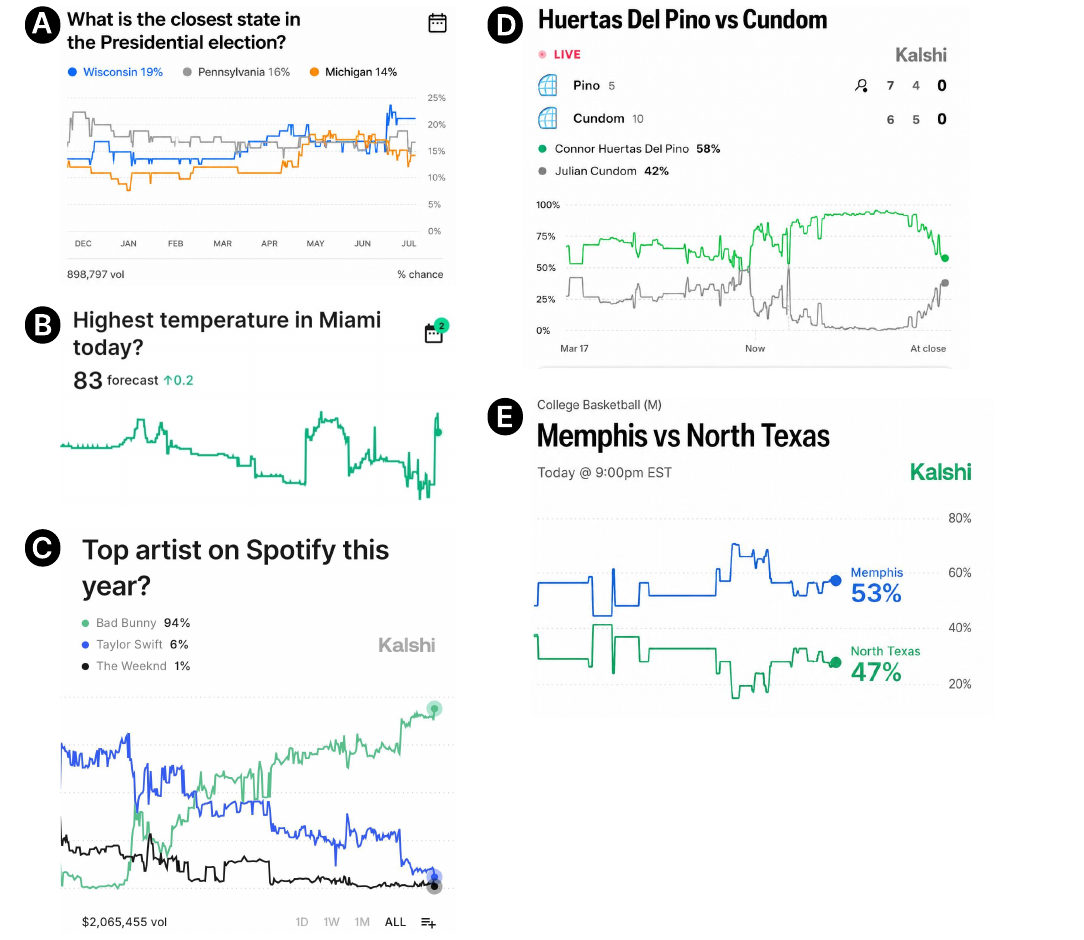}
  \caption{Examples of posts representing the following themes:
  (\textbf{A}) Data Interpretation and Chart Sensemaking,
  (\textbf{B}) Uncertainty and Probability Comprehension,
  (\textbf{C}) External Knowledge and Contextual Reasoning,
  (\textbf{D}) Skepticism and Credibility, and
  (\textbf{E}) Engagement \& Decision-Oriented Reactions.}
  \label{fig:theme-examples}
\end{figure}

Users were trying to understand what the visualization was actually showing.\textit{Data Interpretation and Chart Sensemaking} was the most prevalent theme (n=35), with comments focusing on chart values, trends, prices, payouts, and market information. The most frequent code was chart value interpretation (n=32), followed by market mechanics and payouts (n=10), and tooltips or data-source reading (n=4). 
(Figure~\ref{fig:theme-examples}\textbf{A}) shows example post where users attempted to interpret chart values, trends, and market information.
In these comments, users often struggled to determine what specific visual values represented. Some comments reflected difficulty interpreting the relationship between prices, probabilities, payouts, and outcomes. For example, one user questioned whether buying multiple ``Yes'' shares across related outcomes would function like a combined contract: \textit{``Maybe I misunderstood, but can't you just buy yes shares on Michigan, Pennsylvania, and Wisconsin? ... if any of them hit, you end up with \$1.00."} This suggests uncertainty about how mutually exclusive outcomes and payout logic should be interpreted. In a post a user reflected confusion about whether displayed values should be interpreted as probabilities, market prices, or payouts. In one discussion, a user clarified why percentages did not necessarily add to 100\%: \textit{``The percentages quoted are usually either the last traded price or the midpoint of the current orderbook(s). There is no reason these would add to 100\%, even if proper probabilities do.''} 

These examples suggest that interpretation problems were not limited to chart readability. In several cases, users often had difficulty connecting visualized values to the underlying market mechanism. A probability-over-time chart, price display, or payout table could appear straightforward, but the meaning of the displayed value depended on additional market concepts such as last traded price, order book midpoint, liquidity, contract structure, and settlement rules. In this sense, uncertainty was inferred not only from the event being predicted, but also from ambiguity about what the visualization itself represented.
We also observed comments that showed users relied on visualization elements such as tooltips to clarify meaning. In one example, users discussed whether the chart showed maximum temperature or sampled temperature values. One comment explained: \textit{``That doesn't show max temp. That shows recorded temp at the time it was sampled. Max temp could have occurred between data points.''} Here, the uncertainty came from the relationship between the visualized data and the measurement process behind it. The chart appeared to present a clear value, but users still needed to infer what that value actually represented.

\subsection{\textbf{I don't understand these informations, I need more information : Uncertainty and Probability Comprehension}}

\textit{Uncertainty and Probability Comprehension} refers to comments where users tried to understand what probabilities, forecasts, and uncertain outcomes meant in a prediction market setting. These comments often involved confusion about whether a displayed value represented a probability, a market price, a forecast, or a possible payout. \textit{Uncertainty and Probability Comprehension} appeared in 11 annotated snippets. These comments included probability aggregation or mismatch (n=7), multiple outcomes and hedging logic (n=3), and the distinction between forecasts and market prices (n=2). 
(Figure~\ref{fig:theme-examples}\textbf{B}) presents examples of users questioning probabilities, forecasts, and the meaning of uncertain outcomes.
These comments show that users were not only reading chart values, but also trying to understand how probabilities, prices, and outcomes should be interpreted in a prediction market setting.

One recurring issue was that users expected probabilities to behave like conventional probability estimates, even when the displayed values were tied to market prices. For example, one comment explained that displayed percentages may represent the last traded price or the midpoint of the order book rather than a set of probabilities that should add to 100\%. Another user questioned whether buying shares across multiple related outcomes would create a guaranteed payoff, showing how probability interpretation was tied to contract structure and hedging logic.

We also observed comments that distinguished between a forecast and a prediction market price. In one example, a user stated: \textit{``It's not a forecast, it's people betting on a prediction market.''} This distinction matters because a visualized probability in a prediction market may look like a direct forecast, but it is shaped by trading activity, liquidity, and market participation. These comments suggest that uncertainty was inferred not only from the probability value itself, but also from users' understanding of how that value was produced.

\subsection{\textbf{What I know is more valuable than visualization : \\ External Knowledge and Contextual Reasoning}}

\textit{External Knowledge and Contextual Reasoning} covers posts and comments where users interpreted the visualizations by bringing outside knowledge. These comments drew on sports performance, player injuries, music rankings, political events, weather data, platform rules, and external data sources. \textit{External Knowledge and Contextual Reasoning} appeared in 22 annotated comments. Most comments in this theme involved outside event or domain knowledge (n=18), while a smaller number referred to external sources and resolution rules (n=4). 
(Figure~\ref{fig:theme-examples}\textbf{C}) shows how users brought outside information and prior knowledge into their interpretation of the visualizations. For example, one user explained market movement by referring to entertainment news: \textit{``Taylor Swift was rumored and then Bad Bunny got it which boosted his odds.''} Another user brought in sports knowledge from attending a game: \textit{``I was at the Bears v Green Bay game, and they were down 21-3, and there was talk all around me of how the Bears are a second half team and consistently make great halftime adjustments.''}
In another example, users referred to official data sources or platform-specific rules to explain why a market might resolve differently from what users expected based on common reference points. For example, a comment about a cryptocurrency market explained that the relevant price was not necessarily the one visible on Google or Coinbase: \textit{``Not all cryptocurrency price data is the same. While checking a source like Google or Coinbase may help guide your decision, the price used to determine this market is based on CF Benchmarks' corresponding Real Time Index.''} This kind of comment shows how users used external data-source knowledge to interpret the meaning and reliability of the visualization.

These comments show that prediction market visualizations were rarely treated as self-contained evidence. Users often evaluated the chart in relation to outside events, domain expertise, and assumptions about how a market would resolve. This contextual reasoning could support interpretation, as when users explained why a line moved after new information entered the market. However, it could also produce skepticism or disagreement when users believed that the market price did not reflect what they knew about the event. In both cases, users inferred uncertainty by combining visual information with external knowledge. These patterns are especially important for understanding prediction market visualizations as decision-support systems. A displayed probability or trend did not automatically determine what users believed. Instead, users interpreted it against other sources of evidence. As one user put it, \textit{``Trading based on just odds/money movement without following what's happening in the game is like flying a plane blindfolded.''} For some users, the visualization was useful only when paired with domain knowledge about the underlying event.

\subsection{\textbf{Market is rigged, and misleading : Skepticism and Credibility}}

\textit{Skepticism and Credibility Assessment} theme captures where users questioned whether the visualization and the information could be trusted, especially when values seemed misleading, data sources were unclear, market activity was low, or settlement rules were ambiguous with posts' and comments related to liquidity, trust, or source credibility (n=11), example post (Figure~\ref{fig:theme-examples}\textbf{D}) where users questioned the data, the market, or the platform's representation of information. Users questioned whether the displayed odds reflected true likelihood, whether a market was illiquid or mispriced, whether a source counted as valid for settlement, or whether a chart label accurately represented the underlying data. In one example, a user described a market as ``a gray area'' because a source validated the existence of a speech but did not directly mention the terms needed for market resolution: \textit{``A radio show is not a verified source, but a verified source validated the speech existed on the show itself. BUT, verified source never quoted or mentioned Somalia' or `TDS' in the article. This seems to be a slightly confusing gray area per the official rules."}

Other comments questioned whether the market price reflected meaningful information or was distorted by liquidity. One user wrote: \textit{`` There's no volume in the market probably the only bet that got filled is the 99\% one.''} This comment shows how users evaluated whether a displayed probability reflected broad market belief or only limited trading activity. In another example, a user expressed concern that a visualization was misleading because the displayed percentage reflected one value while the cash-out amount reflected another: \textit{``The UI is confusing/misleading. The 42\% is the last traded price, but the dollar value is how much you'd get if you sold right now.''}


\subsection{\textbf{These charts are not well-designed! : Visualization Design and Usability}}
\textit{Visualization Design and Usability} includes comments where users pointed to design choices that made the visualization harder to interpret. Visualization design choices and usability appeared in 9 annotations. These included comments about visualization clarity, layout, and color (n=7), as well as misleading labels or display mismatches (n=3). Users criticized color choices, confusing layouts, unclear labels, and discrepancies between probability values and payout information. For example, one user criticized the color scheme directly: \textit{``Wtf is this? Pale red and bright red as the two colors for the color scheme? As an app developer myself, these details matter and are sloppy.''} These design concerns mattered because they affected how users inferred uncertainty from the visualization. When users could not tell whether a displayed value represented probability, price, volume, payout, or cash-out value, the visualization became a source of uncertainty rather than a resolution.

\subsection{\textbf{What should I do next buy, hold, or cash out? : \\Engagement \& Decision-Oriented Reactions}}

The \textit{Engagement and Decision-Oriented Reactions} refers to posts and comments where users connected the visualization to possible betting actions. Users often moved from interpreting the visualization to deciding what action it might support (n=17). These comments included advice seeking and betting strategy (n=7), as well as emotional reactions and conviction (n=10). In these cases, users  connected visualized uncertainty to action. Some comments interpreted charts as evidence of opportunity, mispricing, or risk. Users discussed whether a market was worth entering, whether odds were too high or too low, whether an outcome could be hedged, and whether a position could be sold before final resolution. In one example, a user described a team at 47\% not as an underdog but as a market opportunity: \textit{``North Texas is sitting at 47\%, and to me that's not an underdog that's a coiled spring. That's the market mispricing a team that’s been punching above its weight and refusing to break. That green line on the chart? That's not a number. That's a warning to anyone sleeping on UNT. So I'm slamming \$1,000 on North Texas tonight.''}. Other comments connected interpretation problems directly to betting risk. One user warned others not to place a bet without understanding the underlying source: \textit{``Dude don’t make this bet this is the top weekly songs! It is going to be one based off the streams already this week alone.''} Another user framed decision quality around whether a bettor could accurately judge their edge: \textit{``There's nothing necessarily wrong with betting into high prices if you still believe you have an edge. But you have to be able to accurately assess if you have an edge. Most people can't/don't do that.''}



\section{Discussion, Limitations, and Future works}
Our findings  suggest interpreting prediction market visualizations involves more than reading probabilities, prices, or trend lines. Credibility assessment was central to users’ reasoning under uncertainty. Users did not only ask what the chart showed; they asked whether the visualized information should be trusted. Skepticism was often tied to the mismatch between precise looking visual values and the uncertain, conditional, or liquidity dependent nature of the underlying market. A line chart or probability label could make a market appear stable or interpretable, while users’ comments revealed concerns about whether the value was based on meaningful trading volume, a valid data source, or a correct interpretation of the rules.

The findings also show prediction market visualizations are not only interpreted as information, but also as support for action. Users expressed excitement, doubt, frustration, fear, and confidence while reasoning about visualized market information. Some warned others not to bet without understanding the rules or data source, while others treated the visualization as evidence of a profitable opportunity. This reinforces the central point of the study: prediction market visualizations are financially consequential uncertainty visualizations. Even when uncertainty is not explicitly encoded, users infer it from market movement, payout structures, competing outcomes, and visualization details, then connect those interpretations to decisions.

\subsection{Limitation}
This study focuses on visualization presented on kalshi \cite{kalshi2026} and data collected from subreddit r/kalshi \cite{reddit2026}. The findings reflect one platforms community and may not be generalized to other prediction market platforms. Findings are based on public reddit posts and comments, and may not fully capture their reasoning and betting behavior
\subsection{Future work}
Future work could be on testing visualization techniques for representing uncertainty in prediction markets. This includes exploring how visualization could more clearly communicate market price, probabilities, uncertainty, and future studies could also compare uncertainty bands representation, and explanation about visualizations. Another direction is to examine how large lanuguage models (LLMs) could help explain prediction market visualizations, help understand the underlying uncertainty, displayed values, and market movement.

\section{Conclusion}

This study examined how users infer uncertainty from prediction market visualizations presented on kalshi. In our analysis, we observed that users did not always interpret prediction market visualizations as neutral displays of market data. Instead, they use charts and visualization to reason about uncertain future outcomes, evaluate whether the displayed information is trustworthy, and decide whether a market position seems risky, misinformed, or actionable. Across the posts and comments, users moved between visual interpretation, probabilistic reasoning, external contextualization, skepticism, and decision oriented discussion.


\section*{Acknowledgments}
We acknowledge the use of ChatGPT \cite{openai2026chatgpt} to improve the grammar, style, and readability. The tool was used solely for text editing, and played no role in the interpretation or generation of the core findings presented.


\bibliographystyle{abbrv-doi-hyperref}

\bibliography{template}

\appendix 
\crefalias{section}{appendix} 

\end{document}